# Influence ScN protective thin layer on the superconducting properties of ultrathin NbN films

Porokhov N.V.[1], Anikanov A.A.[1], Sirotina A.P.[1], Pershina E.A.[1], Shibalova A.A.[1], Shibalov M.V.[1], Diudbin G.D.[1], Mumlyakov A.M.[1], Trofimov I.V.[1], Vovk N.A.[1], and Tarkhov M.A.[1]

[1]Institute of Nanotechnologies of Microelectronics RAS, Nagatinskaya 16a-11, Moscow, Russia

E-mail: nporokhov@gmail.com

**Abstract**

The present study delves into exploring the impact of a thin protective layer of scandium nitride (ScN) on the superconductive properties of thin films of niobium nitride (NbN) generated through reactive magnetron deposition. This is the first time such an investigation has been carried out. The article offers a comprehensive investigation of the morphological, microstructural, and electrophysical features of thin films that have undergone high-temperature annealing in an oxygen-rich environment. The dependence of the critical transition temperature of the NbN thin film on the annealing temperature of the samples in an oxygen medium, with and without ScN coating, was determined. According to X-ray reflectometry studies, it has been observed that the ScN film serves as a protective layer, even when exposed to annealing temperatures of approximately 450°C, without significantly affecting the NbN layer density or thickness. It has been shown that adding a ScN coating to a thin NbN film increases its resistance to corrosive media compared to a film without the coating.

## 1. Introduction

Niobium nitride is widely adopted in the manufacturing of superconductive nanoelectronic devices of various functionalities, such as single-photon detectors SNSPD [1], hot-electron bolometers (HEB) and mixers of THz frequency range [2], microwave kinetic inductance detectors (MKIDs) [3], etc. Thin NbN films are produced using various techniques, such as reactive magnetron sputtering with Nb target in Ar and $N_2$ gas mixture [4], pulsed laser deposition (PLD) [5], high-temperature chemical vapor deposition (HTCVD), and atomic layer deposition (ALD) [6]. NbN films have been found to exhibit strong compatibility with complementary metal-oxide-semiconductor (CMOS) technology. Their inherent property to react with fluorine renders them highly amenable to processing in industrial plasma chemical etching systems [7]. Throughout the various stages of the technological process involved in the production of superconducting electronics, high temperatures and chemically aggressive atmospheres are often present. This increases the probability of functional layer degradation within components. In this regard, it is important to study the materials of ultra-thin protective layers that do not affect the properties of functional thin films. Scandium nitride exhibits high electronic mobility, hardness (23 GPa), melting point (> 2873 K), and corrosion resistance [8, 9]. Such properties of scandium nitride make it a highly promising material in the development of wear-resistant and protective coatings on various surfaces [10].

The present study investigates the influence of an ultra-thin ScN protective layer, with a thickness of 10 nm, on the electrophysical characteristics of NbN thin films produced via reactive magnetron deposition. A comprehensive investigation of the morphological, microstructural, and electrophysical features of thin films that have undergone high-temperature annealing in an oxygen-rich environment was carried out. The morphology of the obtained films was analyzed using atomic force microscopy (AFM). The films obtained were subjected to structural analysis employing the techniques of transmission and high-resolution transmission electron microscopy, commonly referred to as TEM and HRTEM, respectively. The X-ray reflectometry (XRR) method was the principal technique for regulating the thickness and density of functional films both pre and post annealing. The electrophysical characteristics of the films were evaluated both at ambient and cryogenic temperatures.

## 2. Experimental details

In this study, the samples of two different compositions were analyzed, i. e., "NbN" – Si(n-type) wafer/SiO$_2$/NbN and "NbN/ScN" – Si(n-type) wafer/SiO$_2$/NbN/ScN. The compositions had a single base of a Si(n-type) silicon wafer, which was coated with an amorphous thermal oxide layer, approximately 500 nm in thickness. Subsequently, a thin layer of NbN, with a thickness of 13 nm, and an in-situ ScN thin layer, with a thickness of 10 nm, were deposited ex-situ using a magnetron sputtering method, as illustrated in Figure 1.

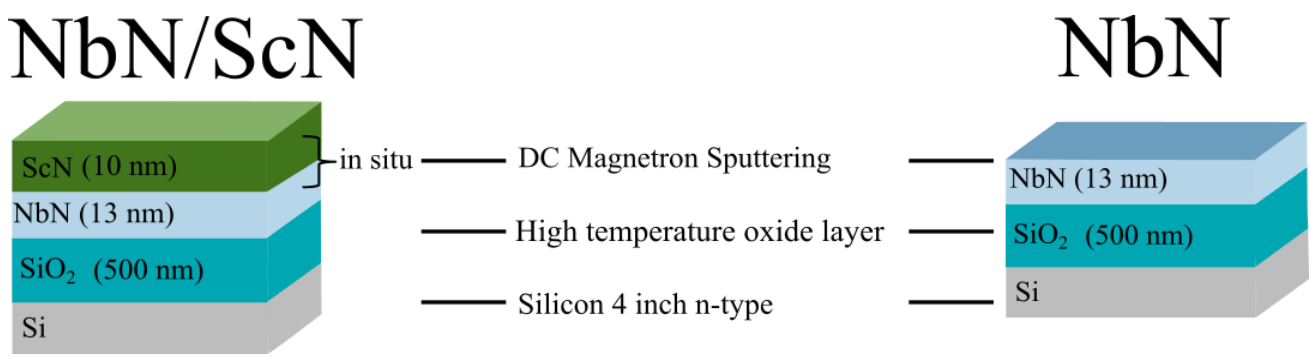

**Figure 1.** Schematic representation of the samples under investigation

The deposition of thin nitride films was achieved through reactive magnetron sputtering from a 99.999% niobium and scandium target in an inert gas mixture of argon and nitrogen at ambient temperature. The ratio of argon and nitrogen was 4:1, respectively. The power of DC magnetron was 5 W/cm$^2$. Under these conditions, the rate of deposition was 11 nm per minute.

After that, the samples were subjected to an accelerated aging process through oxygen annealing at varying temperatures. The annealing process was conducted within a diffusion tube, employing a predetermined thermal program as depicted in Figure 2. The sample was heated to the annealing temperature under a nitrogen environment. This was followed by a temperature stabilization phase that lasted for a duration of 10 minutes. Subsequently, the annealing stage was conducted under dry oxygen for 30 minutes. The heating and cooling rates of the samples were in a ratio of 1:2.

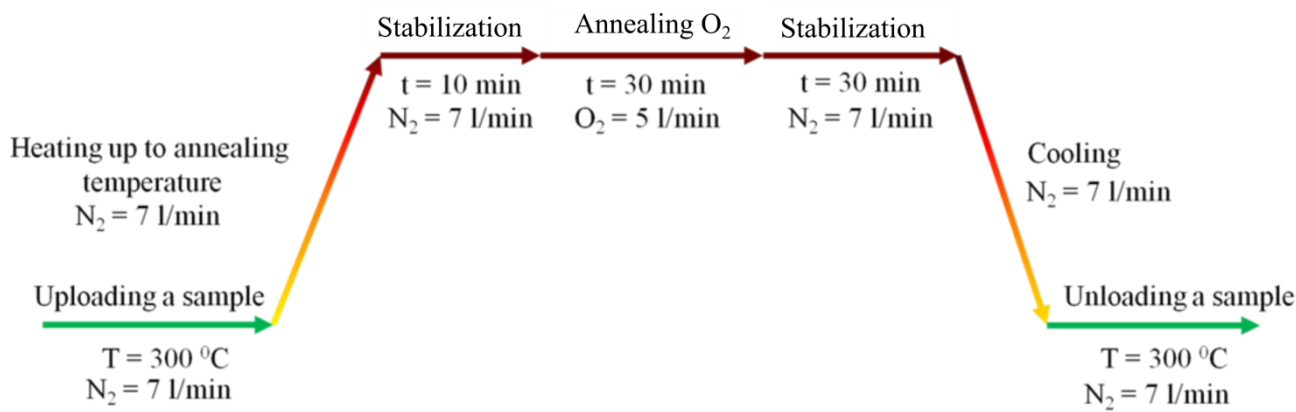

**Figure 2.** Thermal program for oxygen annealing in a diffusion tube under a dry oxygen atmosphere

**Result and discussion**

The surface resistivity ($R_s$) of the film serves as the input parameter for assessing its conductivity. This parameter enables the estimation of the degradation rate of the electrically conductive layer of the sample at varying annealing temperatures. The four-probe method was utilized to determine the surface resistivity value of the films, as stated in reference [11]. The dependence between the annealing temperature and the surface resistivity values for samples of the "NbN/ScN" composition is given in Figure 3, represented by the blue line. The graphical representation reveals that the $R_s$ values remain constant up to a temperature of 450°C, but an increase in resistivity is observed at 550°C. Interestingly, at temperatures of 650°C and beyond, the electrically conductive layer becomes dielectric. The red line in Figure 3 represents the dependence for the "NbN" composition. It is apparent that the deterioration of the electrically conductive layer takes place at lower annealing temperatures compared to the "NbN/ScN" composition.

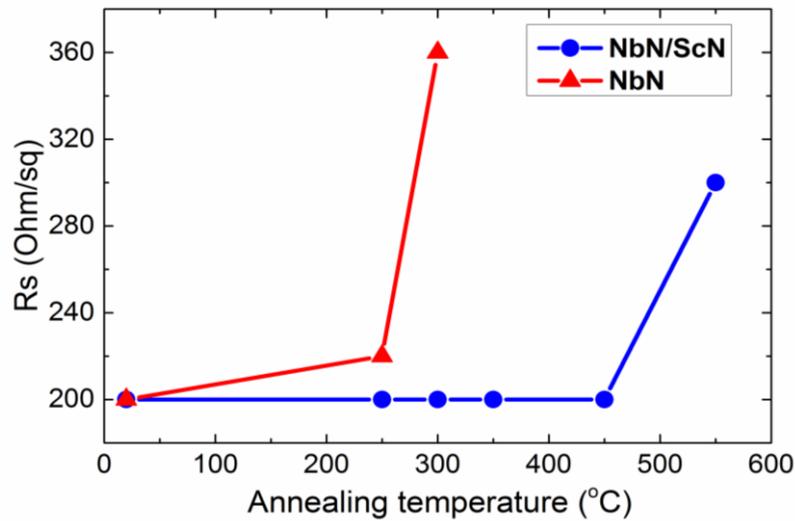

**Figure 3.** Dependence of film surface resistivity (Rs) on oxygen annealing temperature. The red line represents values for the "NbN" composite, while the blue line represents values for the "NbN/ScN" composite

The cryogenic measurements were conducted within a Gifford-McMahon closed-cycle cryostat, where the dependences of the sample resistivity on temperature R(T) were recorded. To calculate the transition critical temperature ($T_c$) value and determine the transition width ($\Delta T_c$), the R(T) dependence diagram was transformed into a dR/dT(T) dependence diagram. In this diagram, the maximum value of the function represents $T_c$, while the width of the diagram at half height corresponds to the width of the sample's transition to the superconductive state. Experimental dependences of the normalized resistivity for samples of "NbN" (a) and "NbN/ScN" (b) compositions are shown in Figure 4 for different annealing temperature conditions.

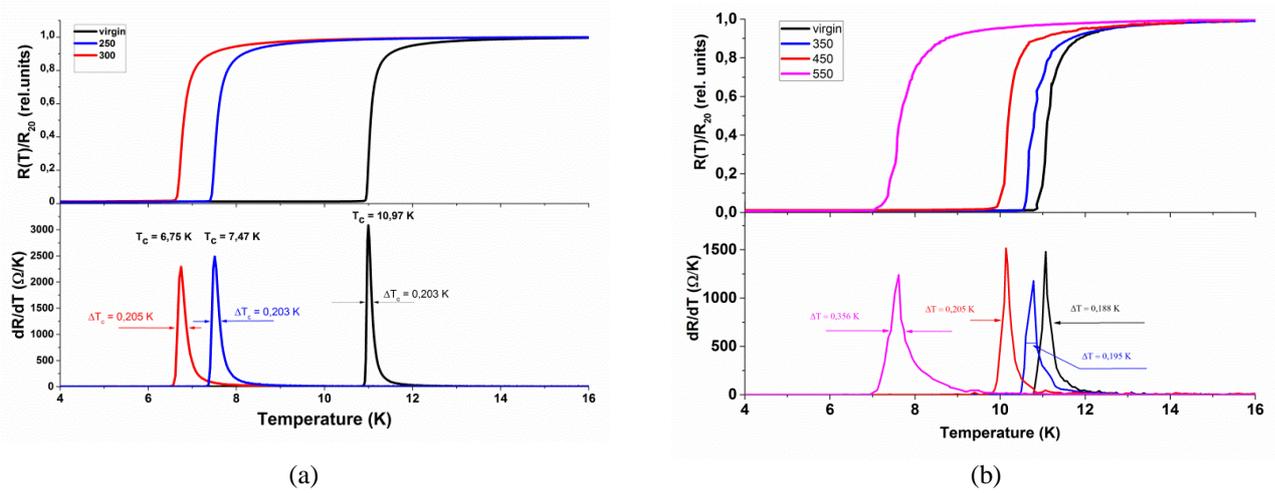

(a)                  (b)

**Figure 4**. Dependence of normalized resistance on temperature: (a) for the "NbN" composition; (b) for the "NbN/ScN" composition subjected to oxygen annealing

As per the anticipated behavior of the "NbN" composition, the critical temperature $T_c$ showed a consistent decrease with a corresponding increase in the annealing temperature, as represented by the red line in Figure 5. At temperatures exceeding 350°C, the "NbN" composition underwent transformation into a dielectric material. The "NbN/ScN" composition exhibits insignificant changes in the critical temperature $T_c$ at annealing temperatures up to 450°C. However, the critical transition temperature experiences a significant shift as the annealing temperature exceeds 450°C, as illustrated by a blue line in Figure 5. The study observed that the ratio of the width of the transition to the superconductive state between the initial samples and samples subjected to almost complete oxidation for both compositions was found to be 1.5 times. It is noteworthy that the width of the transition to the superconductive state of the "NbN/ScN" composition remained nearly constant up to an annealing temperature of 450°C, as depicted in Figure 4 (b). Cryogenic measurements were not conducted for those samples that underwent a dielectric transformation during the annealing stage.

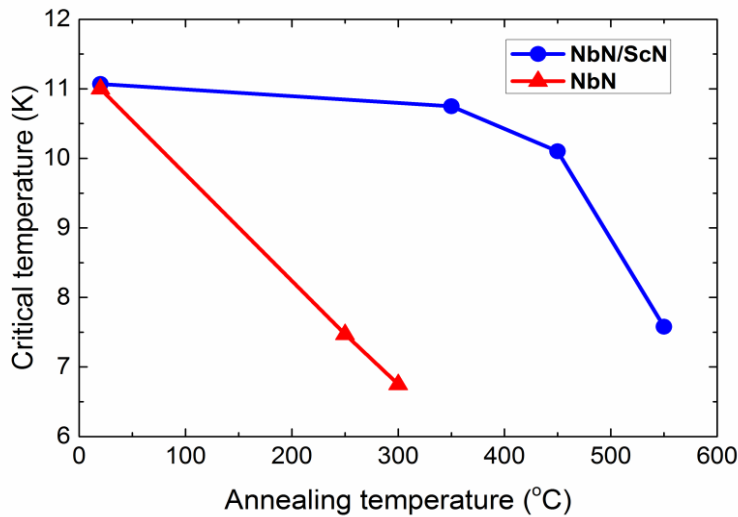

**Figure 5**. Dependence of the critical temperature of samples with compositions "NbN/ScN" and "NbN" on the annealing temperature in an oxygen atmosphere

The critical current measurements were carried out as per the two-point measuring diagram, employing a Keithley 6221A high-precision low noise current source. The dependence between the critical current density and temperature was analyzed based on the current-voltage curves that were obtained during the process of constant-voltage regulation. The current-voltage curves (CVCs) were measured under constant-voltage conditions in the temperature range from 2.5 K up to the resistive transition temperature. The value of the critical current $I_c$ was defined as the current corresponding to a complete structure transition from a superconductive to a resistive state. The dependences of the critical current density on temperature were determined for the "NbN/ScN" composition at different annealing temperatures. The critical current density $J_c(T)$ was determined as $I_c(T)/w*d$, where w is the width of a superconductive strip, and d=13 nm is the NbN film thickness. In Figure 6 (a), the set of temperature dependences $J_c$ is presented. Solid lines represent the predicted dependence of the critical steaming current, which is described in the Ginsburg-Landau theory [12]. Figure 6 (b) illustrates the dependence between the critical current density at liquid helium temperature of 4.2 K and the annealing temperature. Similar to the critical temperature, the diagram of the critical current density's dependence on the annealing temperature exhibits comparable behavior. It has been observed that temperatures exceeding 450°C can lead to noteworthy variations in the critical current density values.

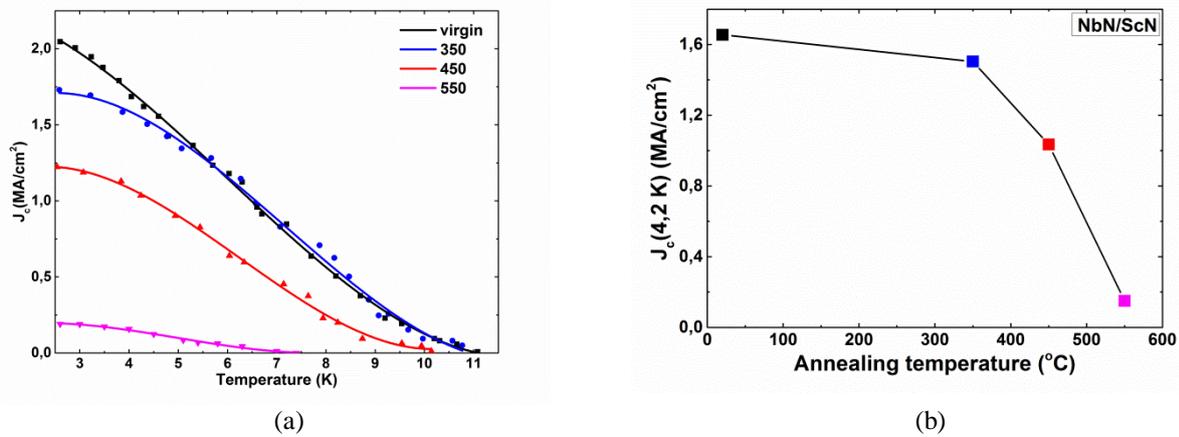

(a) (b)

**Figure 6**. (a) Dependence of critical current density on temperature for the "NbN/ScN" composition at various annealing temperatures; (b) Dependence of the critical current density value at 4.2 K on the annealing temperature of samples with the "NbN/ScN" composition

The present study involved the investigation of surface morphology of films before and after oxygen annealing. The atomic force microscopy (AFM) method was employed to determine the surface roughness of the films. The film surface roughness was analyzed on an area of 1 μm² in a semi-contact mode. After processing the results, parameters of roughness for defined areas were obtained, where $R_q$ is the root mean square deviation of the surface height from the weighted average value. The AFM images of the "NbN/ScN" composition were taken in semi-contact mode for films without annealing and films annealed at 350°C (b), 450°C (c), 550°C (d), 650°C (e), and 750°C (f), respectively.

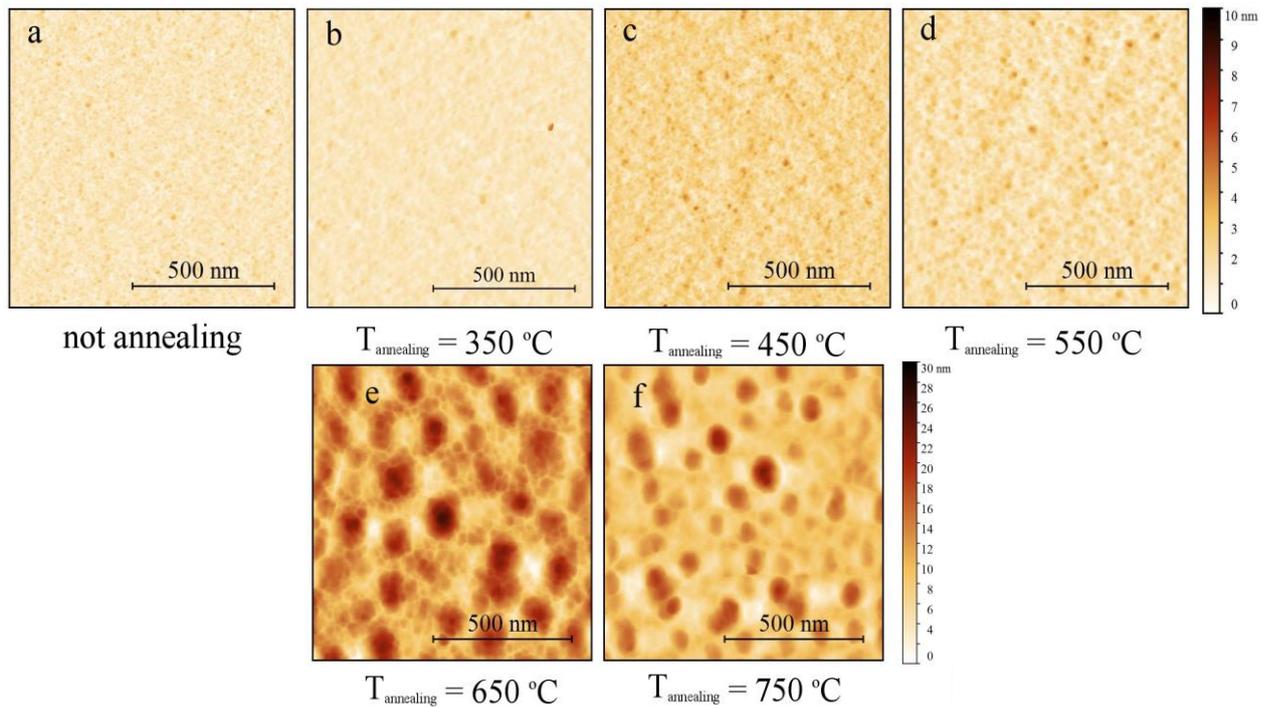

**Figure 7.** AFM images of the surface of film samples with the "NbN/ScN" composition at various annealing temperatures: (a) without annealing, (b) annealed at T = 350°C, (c) annealed at T = 450°C, (d) annealed at T = 550°C, (e) annealed at T = 650°C, (f) annealed at T = 750°C

The AFM images prove that the surface morphology of the ScN film remains relatively stable up to a temperature of 550°C. However, beyond this temperature, the surface morphology undergoes a significant transformation, leading to the formation of large growths at 650°C. Samples of the "NbN" composition exhibit a change in morphology at significantly low temperatures. To illustrate, at 350°C, a considerable number of growths are observed on the surface, as depicted in Figure 8.

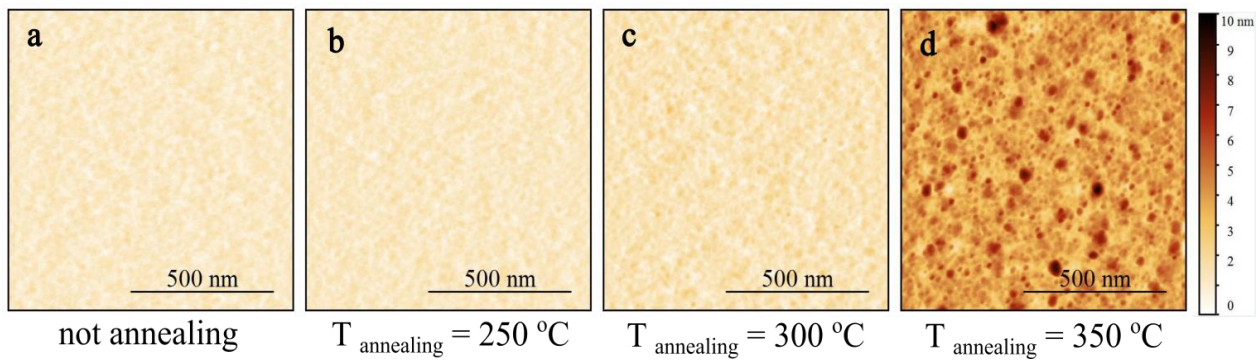

**Figure 8.** AFM images of the surface of film samples with the composition "NbN" at various annealing temperatures: (a) without annealing, (b) annealed at T = 250°C, (c) annealed at T = 300°C, (d) annealed at T = 350°C

The X-ray reflectometry (XRR) method was the principal technique for regulating the thickness and density of functional films both pre and post annealing. The experimental results of X-ray reflectometry for samples of the "NbN" (a) and "NbN/ScN" (b) compositions, along with the initial sample, and samples subjected to oxygen annealing in the temperature range of 250 to 750°C for a duration of 30 minutes are presented in Figure 9.

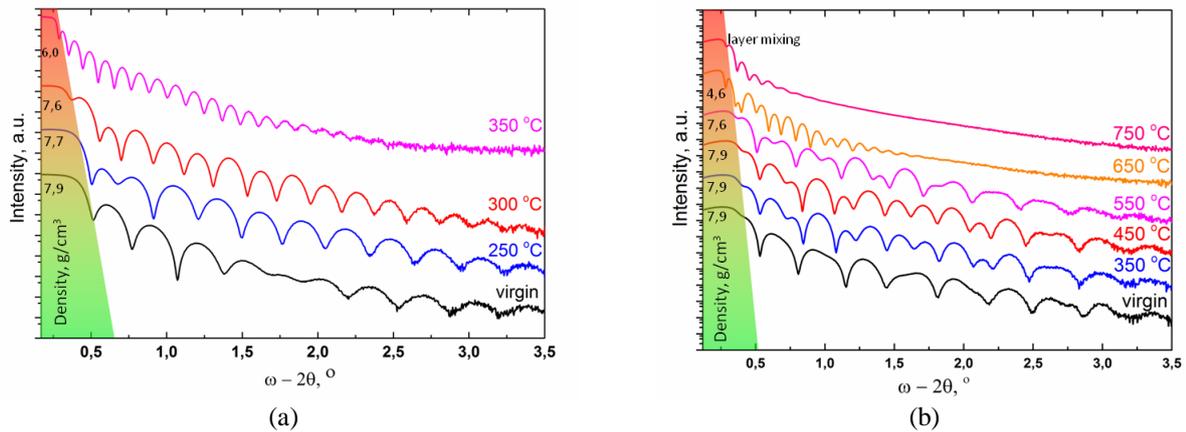

**Figure 9.** Curves of X-ray reflectometry: (a) for the "NbN" composition; (b) for the "NbN/ScN" composition at various annealing temperatures, with inserted values of the measured density of the functional film represented in color gradient

According to the analysis of X-ray reflectometry curves conducted for the "NbN" composition, it has been observed that upon annealing at a temperature of 350°C, the functional layer of NbN undergoes complete oxidation, leading to the formation of a less dense $Nb_xO_y$ material. The results of the studies conducted on the "NbN" composition reveal that the functional film undergoes degradation when exposed to annealing temperatures of 350°C in an oxygen environment. According to the findings of X-ray reflectometry conducted on the "NbN/ScN" composition, the ultrathin ScN film serves as a protective layer at temperatures up to 450°C in an oxygen environment and does not impact the density and thickness of the functional NbN layer. This observation is supported by the transmission electron microscopy (TEM) analysis depicted in Figure 10.

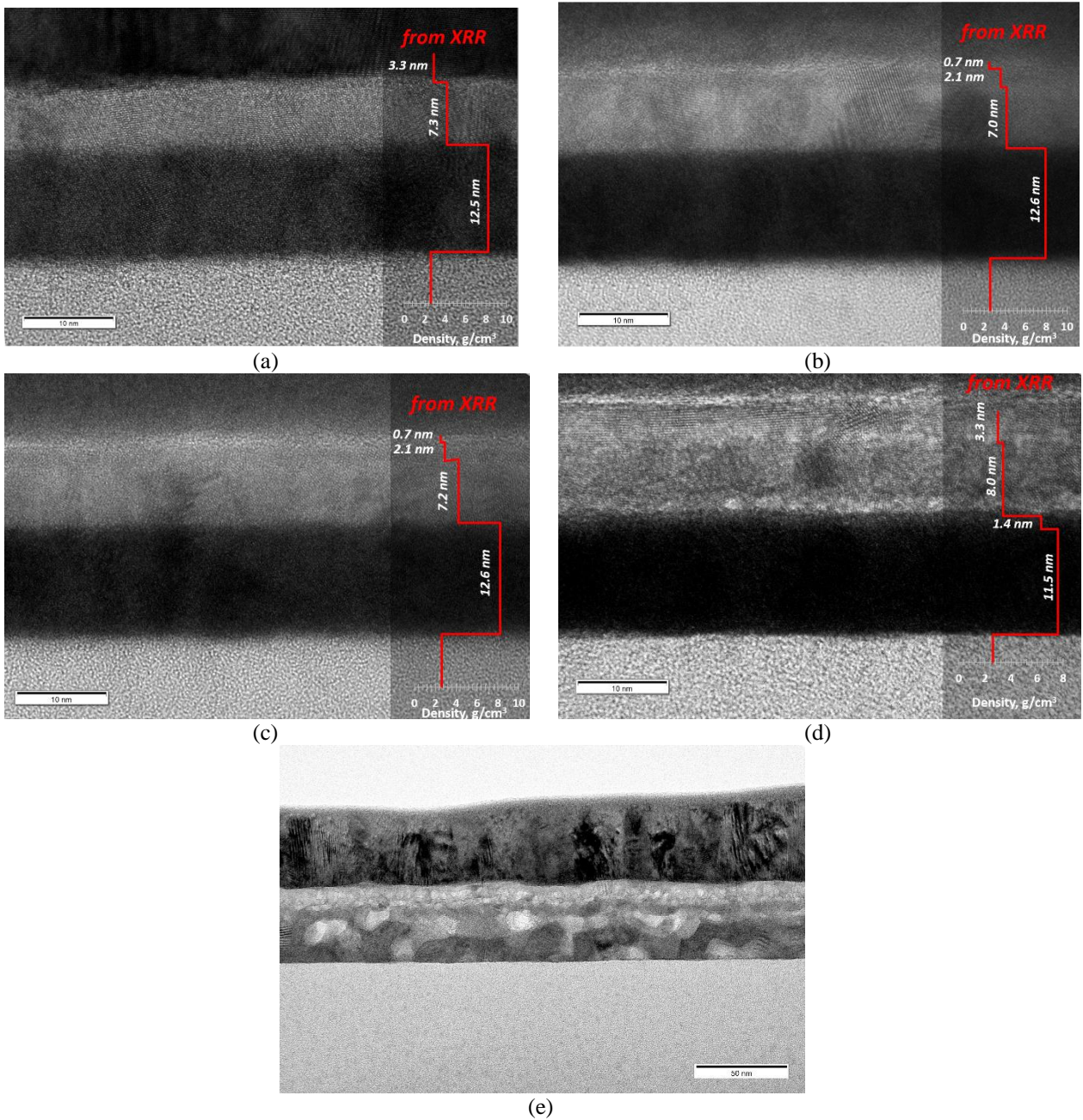

**Figure 10**. TEM images of cross-sectional views: (a) of the initial "NbN/ScN" composition and samples subjected to oxygen annealing at temperatures of: (b) 350°C, (c) 450°C, (d) 550°C, (e) 650°C

TEM results for the initial sample and samples annealed in oxygen at 350°C, 450°C, 550°C, and 650°C are presented in Figure 10. The structure comprises of several layers, where the lowermost layer is composed of silicon oxide. The subsequent dark layer with a high density consists of $NbN_x$, having a thickness of approximately 12.5 nm. Following this, the light layer with a low density is formed by $ScN_x$, having a thickness of around 7.5 nm. The surface layer is less than 3 nm, and the topmost layer is composed of platinum, serving as a protective or contrast layer. On the right of each TEM image, a diagram of the density dependence on the depth obtained from X-ray reflectometry data is superimposed. The comparison of the TEM and XRR data reveals a conspicuous degree of data convergence. According to the observations made using TEM and XRR, it has been found that the ScN protective layer undergoes some changes after annealing at a temperature of 550 °C. Specifically, the layer oxidation occurs. It is worth noting that a thin layer with a thickness of approximately 1.5 nm is generated at the ScN/NbN interface, and a reduction of roughly 1 nm in the NbN overall thickness can be observed. Thus, it can be inferred that the $NbN_x$ layer undergoes partial oxidation at a temperature of 550 °C. At the same time, the sample annealed at 650 °C shows strong changes in both the ScN and NbN layers. The structure of both layers is heterogeneous: a large number of voids, regions of reduced density and different crystal structures are observed along the thickness of each layer. Figure 10e also shows that the roughness of the ScN/NbN interface has strongly increased, the separation into layers is still present and is visually possible by the contrast of the NbN and ScN layers in the TEM images. According to the TEM and XRR data, the thickness of the ScN and NbN layers after annealing at 650 °C has increased significantly and according to the HRTEM results is ≈ 11 nm and ≈ 31 nm respectively compared to ≈ 12.5 nm and ≈ 7.5 nm in the original sample and in the samples after annealing up to 550 °C.

Based on the experimental data set forth above, it is evident that the 10 nm thick ScN thin film exhibits remarkable thermal and chemical stability, thus rendering it a highly promising thin-film material for deployment in high-temperature environments and aggressive chemical settings. However, it remains unclear which factors influence the properties of samples during the annealing process. To obtain a more comprehensive insight into the matter at hand, we have carried out a set of supplementary experiments pertaining to the annealing process of the "NbN/ScN" composition in a nitrogen-rich environment at a similar temperature range (refer to Figure 2), with the exception of oxygen.

The diagram presented in Figure 11 depicts the comparative dependences of the normalized resistance on the temperature of samples of the "NbN/ScN" composition, which were subjected to heat treatment in nitrogen and oxygen media in the temperature range from 20°C to 550°C. Analysis of the diagram suggests that changes in the superconductive properties of the samples are directly related to the influence of external oxygen.

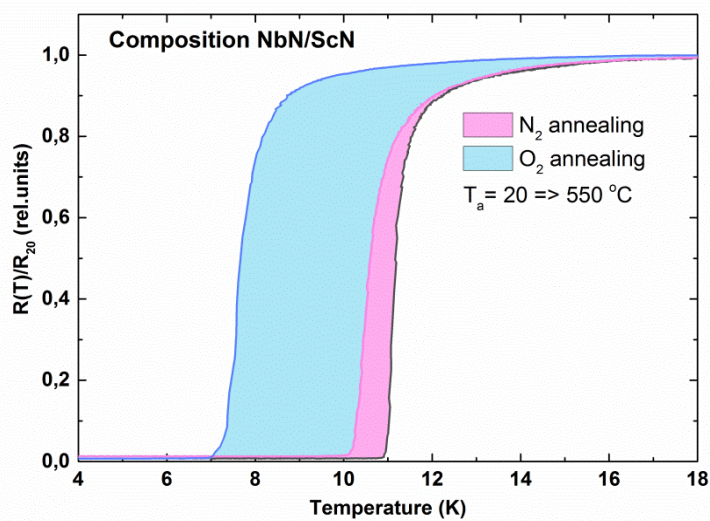

**Figure 11**. Comparative diagram of dependencies of normalized resistance on temperature for samples of the "NbN/ScN" composition, subjected to thermal treatment in nitrogen (pink color) and oxygen (blue color) environments

**Conclusion**

In the course of investigating the protective effect of the ScN thin film, our research revealed that a thickness of 10 nm confers exceptional thermal and chemical stability. Such properties are attributed to the distinct crystal structure of scandium nitride, which imparts high resistance to external factors to the film. According to X-ray reflectometry studies, it has been observed that the ScN film serves as a protective layer, even when exposed to annealing temperatures of approximately 450°C, without significantly affecting the NbN layer density or thickness. It has been demonstrated that adding a ScN coating to a thin NbN film increases its resistance to corrosive media compared to a film without the coating. These observations provide a promising outlook for the applicability of the material under conditions of high temperatures and corrosive chemical environments.

**Acknowledgment**

The authors would like to thank P.A. Nekludova and E.M. Eganova for their assistance in preparing samples for TEM, and Yu.P. Korneeva for preparing samples and performing cryo-measurements.

The study has been executed with the support of project No. 122040800153-0 of the Ministry of Education and Science of the Russian Federation. Fabrication and characterization were carried out at large scale facility complex for heterogeneous integration technologies and silicon +carbon nanotechnologies.